\documentstyle[twoside,fleqn,espcrc2,epsf]{article}

\input{epsf}

\title{Dynamics and Melting of Stripes, Crystals, and Bubbles
with Quenched Disorder}
\author{C. J. Olson Reichhardt, C. Reichhardt, I. Martin, and A.R. Bishop
\address{Center for Nonlinear Science and Theoretical Division, 
Los Alamos National
Laboratory, Los Alamos, NM 87545.}
}
\begin{document}

\begin{abstract}
Two-dimensional systems in which there is a competition between long-range 
repulsion and short range attraction exhibit a remarkable variety of patterns 
such as stripes, bubbles, and labyrinths.  Such systems include magnetic films,
Langmuir monolayers, polymers, gels, water-oil mixtures, and two-dimensional 
electron systems.  In many of these systems quenched disorder from the 
underlying substrate may be present.  We examine the dynamics and stripe 
formation in the presence of both an applied dc drive and quenched disorder. 
When the disorder strength exceeds a critical value, an applied dc drive can 
induce a dynamical stripe ordering transition to a state that is more ordered 
than the originating undriven, unpinned pattern. 

\vspace{1pc}
\end{abstract}

\maketitle

The growing interest in nonextensive thermodynamics and statistical
mechanics represented in these Proceedings is stimulated by
observations (from real experiments or simulations of models) of
phenomena such as incomplete (e.g., fractal) coverage of phase space,
rare-event statistics, intermittency, stochastic resonance, and
so forth.  We can hope for, and certainly seek to find, 
underpinning mechanisms and classifications of such systems.

We suggest that among the relevant classifications are (a) systems
exhibiting mesoscopic self-assembly (usually from nonlinearity or
competing scales at more microscopic scales), and (b) systems with
long-range interactions (and often coexisting short-range interactions
and entropic forces).  Long-range interactions may be in the 
{\it microscopic} variables of the system (e.g., Coulomb fields), 
or they may be induced
at {\it mesoscopic} scales because of local constraints or symmetries
(e.g., vortices in fluids and magnets, patterns in convection cells
and vibrated granular materials, strain-strain interactions in
elastic media).  It is important to distinguish topological (singularity)
mesoscale excitations, which have long-range mutual interactions (such
as vortices, dislocations, stress filaments) from mesoscopic collective
excitations with short-range interactions (such as solitons, breathers).
We expect that the former will exhibit mesoscopic statistics driven by 
their mutual interactions, whereas the latter should show mesoscopic
statistics driven by the (colored, multiplicative) noise bath of the
non-collective degrees of freedom.

We have studied aspects of anomalous statistics and dynamics in a
number of physically motivated models, including: soliton
diffusion \cite{Rasmussen}; relaxation in two and three dimensional complex
Ginzburg-Landau equations \cite{Aranson}; multiscale structure and
dynamics in elastic media \cite{Shenoy}; filamentary transport
in superconductor flux lattices \cite{VortexRiver}; surface morphology
and growth \cite{Sanchez}; relaxation in magnets with planar spin
anisotropy \cite{Komineas}; glassy dynamics near a spinodal in a
Van der Waals fluid \cite{Loh}; charge, spin ordering and dynamics
in strongly correlated electronic materials \cite{Branko}; and large-scale
materials deformation (fracture, friction) \cite{Roeder}.

Motivated by the similarities in the phenomena exhibited by these
(and related) model systems, we have begun studies of a ``minimal'' model
with coexisting short- and long-range interactions.  We introduce this
model here, and briefly report some of its properties.

Recent proposals for ordering in 
two-dimensional (2D) electron systems consider the
idea of an electron liquid crystal system \cite{Fogler,Kivelson,Fradkin}.
Such a system can form stripe phases with varying degrees of disorder:
An anisotropic Wigner crystal is well-ordered in both directions;  a
smectic phase has ordering along only one direction;  in a nematic,
dislocations are introduced into the smectic phase but there is still
an overall preferred orientation for the system;  finally, in an isotropic
liquid stripe phase, the stripe pattern is filled with dislocations and
has no unique orientational direction.

Stripe patterns are very general and are formed in a wide range of
systems \cite{Seul}.  
For example, the Swift-Hohenberg model 
produces glassy
stripe configurations \cite{Boyer}.  
Other stripe-forming systems include magnetic films \cite{Wolfe}, 
Langmuir monolayers,
polymers, gels, and water-oil mixtures \cite{Gelbart}.
One system in which stripes
have attracted a particularly large amount of interest is holes
in layered transition metal oxides.  In a quasiclassical
description,
the interaction between two holes consists of both a long-range
repulsion generated by the Coulomb interaction, as well as a short-range
attraction that arises due to the breaking of the antiferromagnetic
bonds surrounding the holes.  As a result, the holes can form linear
arrays termed stripes.  There is experimental evidence for the presence
of stripes in several of the lanthanum cuprate oxides, but it is not yet
clear whether the high-Tc superconducting compounds such as YBCO
contain stripes, nor what the role of the stripes would be if they
exist.  Experimental detection of stripes is made more difficult by
the possible interaction of impurities in the material with the stripes.
The effect of disorder on a stripe phase is not well understood.  Therefore
we have constructed a simple model containing the basic physics of the
system in order to gain a better understanding of the ordering and
dynamics of a stripe phase.

Our simple model consists of particles with competing long-range and
short-range interactions.  The particles experience a long-range repulsion
of a Coulomb form, and a short-range attraction which we assume to be
a simple exponential form.  When these two interactions are combined, a
new length scale emerges which depends on the density of the particles and
the strength of the short-range attractive term.  We further add to
our model quenched disorder to represent impurities in the material, as
well as an external field, which would be an electric field 
in the case of holes.

The particles are assumed to obey overdamped Langevin dynamics given by
\begin{equation}
{\bf f}_i = -\sum_{j=1}^{N_v} \nabla U(\rho) + {\bf f}^T + {\bf f}^p + {\bf f}^d = \eta {\bf v}_i.
\end{equation}
The potential between two particles is
\begin{equation}
U(\rho)=\frac{q}{\rho} - B \exp(-\rho/\xi) ,
\end{equation}
where $q=1$ is the charge on each particle, and $\xi$ is the magnetic
screening length.  The long range interactions are treated
as in Ref. \cite{Jensen}.
Temperature is represented by random thermal
kicks which have the property,
$<f(t)^T>=0$, $<f(t)_i^Tf(t^\prime)_j^T>=2\eta k_B T \delta_{ij}\delta(t-t^\prime)$.
The disorder is modeled as random, point-like pins, represented by
parabolic traps.  The pinning force is given by
\begin{equation}
{\bf f}_i^p=\sum_{k=1}^{N_p}\frac{f_p}{\xi_p}|{\bf r}_i-{\bf r}_k^{(p)}|
\Theta\left(\xi_p - |{\bf r}_i - {\bf r}_k^{(p)}\right)
\hat{\bf r}_{i,k} ,
\end{equation}
where $f_p=0.2$ is the pinning force, 
$\xi_p=0.125\xi$ is the pin radius, and $\Theta$
is the Heaviside step function.  The force from an applied 
field (voltage) is modeled
as a uniform drive, ${\bf f}_d=f_d \hat{x}$.
We measure the velocity signature $<V_x>$ of the system, which would correspond
to a current. 

\begin{figure}
\center{
\epsfxsize=3.2in
\epsfbox{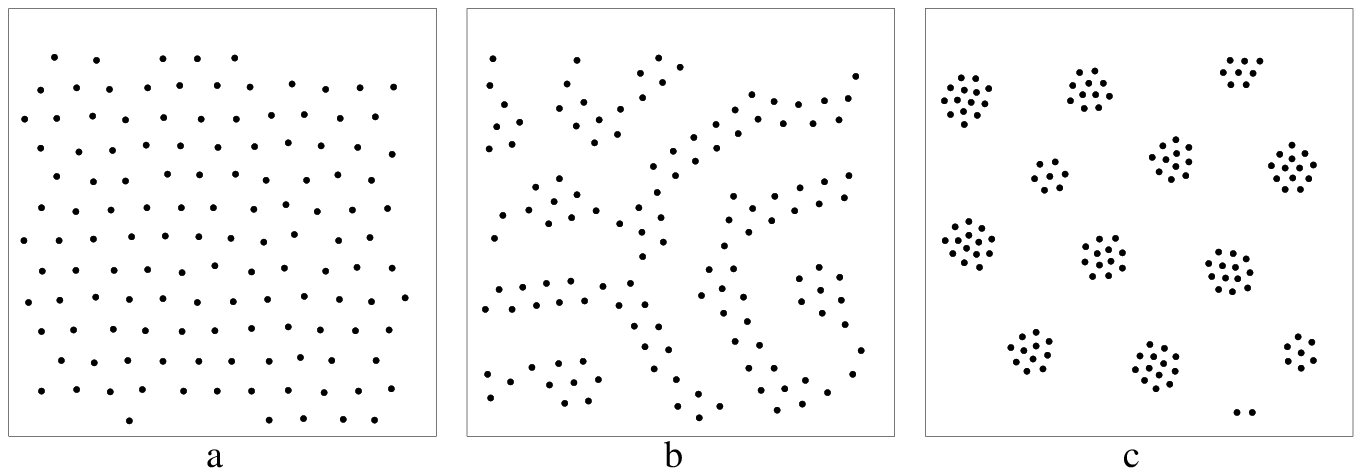}}
\caption{
Static phases as a function of increasing strength of the
short range attraction, $B$. (a) Wigner crystal phase at low $B$. 
(b) Stripe phase at intermediate $B$. (c) Clump phase at high $B$.
}
\label{fig:static}
\end{figure}

\begin{figure}
\center{
\epsfxsize=3.2in
\epsfbox{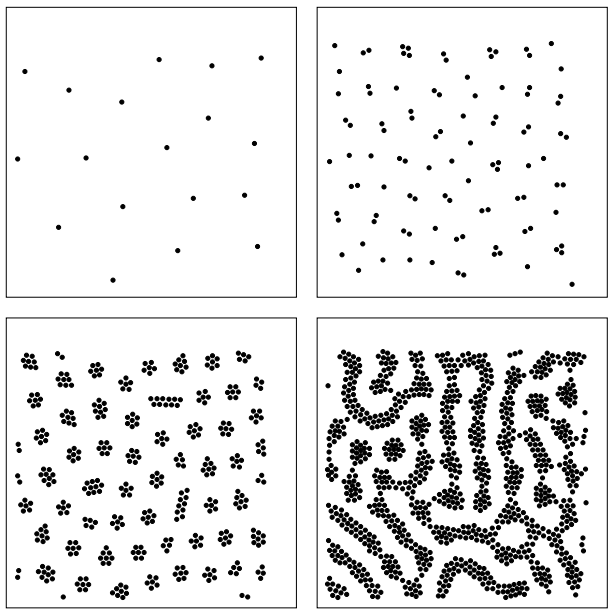}}
\caption{
Static phases 
for increasing particle density
with fixed $B$. 
}
\label{fig:static2}
\end{figure}

We initially consider a system with no pinning or external drive.  We
vary the strength of the short range attraction by changing the value
of $B$, while holding the particle density fixed at $n=0.64/\xi^2$  
As shown in Fig.~\ref{fig:static}(a), when $B$ is small, the long
range repulsion dominates and the system forms a Wigner crystal.
At high values of $B$, Fig.~\ref{fig:static}(c),
the short range attraction dominates and the
particles form clumps which become larger as the value of $B$ is further
increased.  The clumps organize into a superstructure of a triangular
lattice of clumps.
At intermediate values of $B$, Fig.~\ref{fig:static}(b),
the short and long range
interactions are of comparable strength, and the system forms a
disordered stripe structure.  We observe the same phases if we
fix the value of B and vary the density of the system.  This
is illustrated in Fig.~\ref{fig:static2}.

\begin{figure}
\center{
\epsfxsize=3.2in
\epsfbox{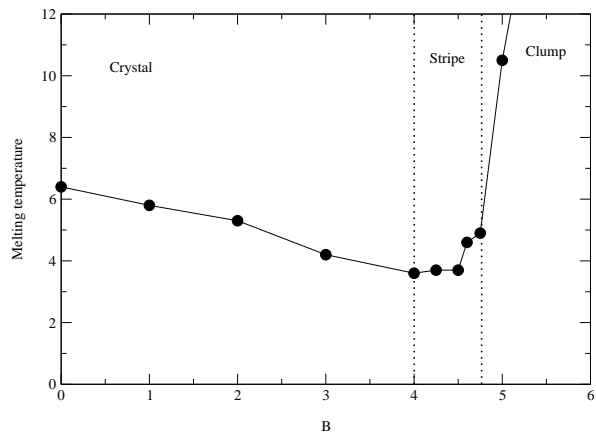}}
\caption{
Melting temperature for the three phases as a function of $B$.
}
\label{fig:melt}
\end{figure}

\begin{figure}
\center{
\epsfxsize=3.2in
\epsfbox{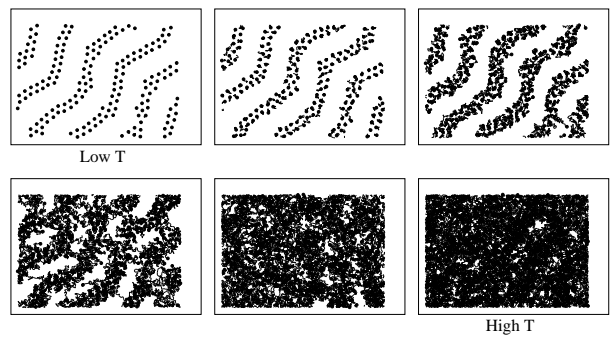}}
\caption{
Snapshots at different temperatures
illustrating the melting of the stripe phase.  Black dots are the
particles; lines indicate the trajectories of the particles over a fixed
interval of time which is the same for all panels.
}
\label{fig:stripemelt}
\end{figure}

We have measured the melting temperature of the three phases by
computing the particle diffusion at each temperature.  We define
the melting temperature to be the temperature at which the particles
first diffuse more than a lattice constant.  The melting temperature
$T_m$ as a function of $B$ is plotted in Fig.~\ref{fig:melt}.  
$T_m$ decreases in the crystal phase as $B$ increases since the
short range attractive term introduces distortions in the crystal,
allowing it to melt more easily.  The stripe phase has the lowest melting
temperature, but $T_m$ increases dramatically upon entering the
clump phase.  This is because the clump phase contains a large
amount of elastic energy.  Within each clump there is triangular
ordering of the particles, and additionally the clumps form an
ordered superstructure.  We find that the clump phase simultaneously
melts and dissociates for the parameters considered here.
In contrast, in the stripe phase melting first occurs along the length
of each stripe, as illustrated in Fig.~\ref{fig:stripemelt}.  Particles
diffuse freely within the stripes but do not enter the regions between
the stripes until higher temperatures are applied.  Thus we find
evidence for a two-stage melting of the stripe phase.

\begin{figure}
\center{
\epsfxsize=3.2in
\epsfbox{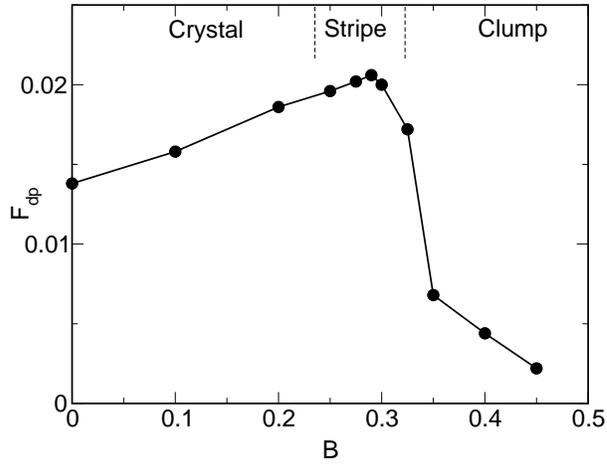}}
\caption{
Depinning force $F_{dp}$ for the three phases as a function of $B$.
}
\label{fig:depin}
\end{figure}

We next consider the effect of disorder on the three phases.  We 
add random point-like disorder to the system, and then apply a
driving force to the particles.  At small drive strengths, the particles
are trapped by the disorder and do not move.  We measure the threshold
force required to depin the particles in each phase.  As shown
in Fig.~\ref{fig:depin}, the depinning force increases with
increasing B in the crystal phase, is highest in the stripe state, and
then drops dramatically upon entering the clump phase.  This is
the same behavior, but inverted, that was observed for the melting
transition in Fig.~\ref{fig:melt}.  This similarity is due to the fact
that the softness of the stripe phase, which allows it to melt at lower
temperature, also allows it to be well pinned by the randomly located
pinning sites.  The stripe state is disordered and can readily readjust
to take advantage of the maximum number of pinning sites.  In contrast, in
the highly elastic and stiff clump phase, the particles are so constrained
by the ordering of the phase that it is very difficult for them to adjust
in order to find the lowest energy pinning location.  As a result many
of the pins are ineffective in the clump phase and the depinning force is
low.  This similarity between the melting and depinning signatures is
particularly interesting as it implies that information about a thermal
transition (melting) can be obtained from a dynamical experiment at constant
temperature (depinning).

\begin{figure}
\center{
\epsfxsize=3.2in
\epsfbox{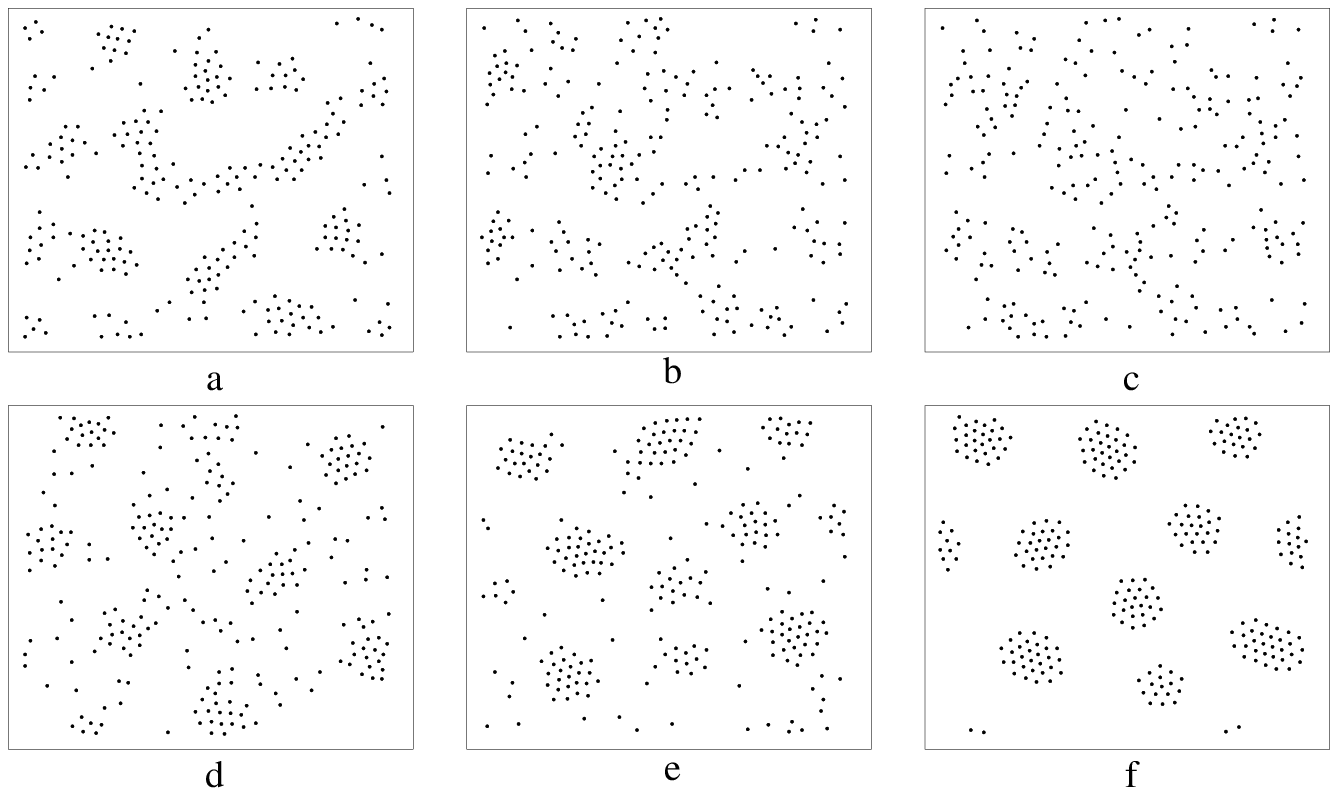}}
\caption{
Dynamic reordering of the clump phase.  
The external driving
force increases from panel (a) to panel (f).
}
\label{fig:clumporder}
\end{figure}

At high applied drives, it is possible to (dynamically) 
reorder the phases, which are
distorted by the presence of the pinning at low drives.  Dynamical 
reordering of a triangular lattice of repulsively interacting particles
has been observed previously in several systems, including superconducting
vortices \cite{Vinokur,VortexReorder} 
and Wigner crystals \cite{WignerReorder}.
At high drives, the force from the pinning begins to resemble a 
temperature which decreases as the drive increases, until at infinite
drive the effective temperature is zero, leading to the concept of
dynamical freezing \cite{Vinokur}.  
Dynamical freezing has already been observed in
the Wigner crystal phase.  We find that the clump phase also undergoes
a dynamical freezing transition, and that at high drives the clump phase
in the presence of disorder has the same form as the static clump phase
in the absence of disorder, as shown in Fig.~\ref{fig:clumporder}.

\begin{figure}
\center{
\epsfxsize=3.2in
\epsfbox{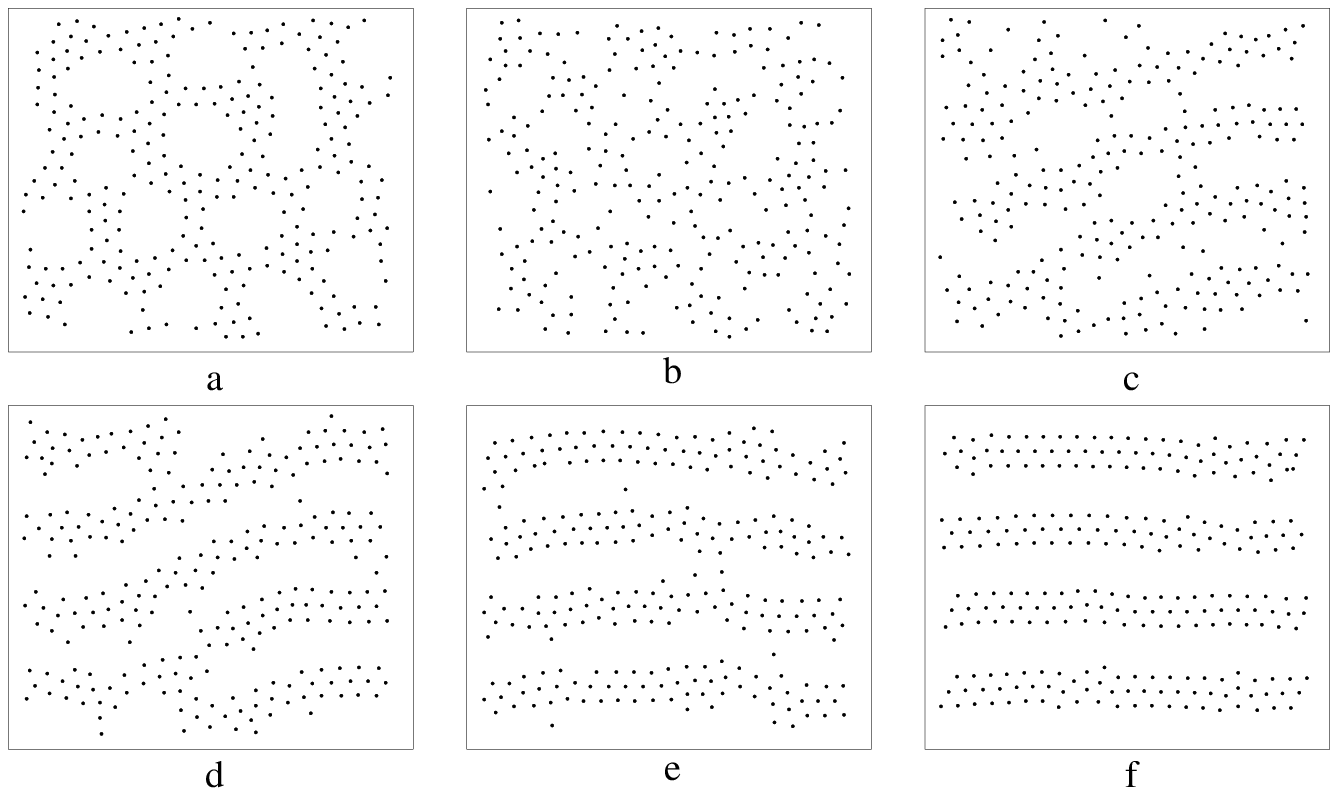}}
\caption{
Dynamic reordering of the stripe phase.
The external driving
force increases from panel (a) to panel (f).
}
\label{fig:stripeorder}
\end{figure}

When we consider the dynamical reordering of the stripe phase, illustrated
in Fig.~\ref{fig:stripeorder}, we find that the final ordered stripe phase
is {\it more ordered} than the corresponding stripe phase that appears
in a static system without pinning.  The stripes are {\it aligned} due
to the driving force, rather than being randomly aligned as was the
case in the thermally annealed system.  This indicates that, in a stripe
forming system, it may be possible to use a combination of quenched
disorder and an external drive to produce well-ordered stripes.  The
reordering of the stripe phase occurs only if the disorder is 
sufficiently strong to cause the stripe to undergo disordered, plastic
flow at low drives.  The original ordering of the stripe must be
destroyed by the disorder in order to allow for a new ordering of
the stripe to form aligned with the drive.  As a result of the 
reordering of the stripe phase, only this shows hysteresis
if the driving force is increased and then decreased again.  The
other two phases, clump and Wigner, do not exhibit any hysteresis.

In summary, the competition between short range attraction and long
range repulsion leads to the formation of Wigner crystal, stripe, and
clump phases as a function of either the strength of the short range
term or the particle density.  The melting temperature is lowest in
the stripe phase, while the depinning force is highest in the stripe
phase, and lower in the Wigner and clump phases.  Quenched disorder
distorts the ordered phases, but the structure of all three phases
can be regained through dynamical reordering at high drives.  The
dynamically reordered stripe phase is more ordered than the equilibrium
stripe phase in a clean system, and as a result the stripe phases
exhibits hysteretic signatures which could be observed experimentally.

In the context of ``nonextensive thermodynamics,'' it is now clearly
of interest to study the various mesoscopic phases of our model (and
their dynamics), and to compare their statistics and responses with
predictions of Tsallis or other superstatistics.

This work was supported by the US Department of Energy under contract
W-7405-ENG-36.

\end{document}